\documentclass[10pt]{iopart} 

\usepackage{harvard}
\expandafter\let\csname equation*\endcsname\relax
\expandafter\let\csname endequation*\endcsname\relax
\usepackage{graphicx,amssymb,amsmath}

\makeatletter
\g@addto@macro\@floatboxreset\centering
\makeatother

\begin{document}

\title[Ultra-fast escape maneuver of an octopus-inspired robot]{Ultra-fast escape maneuver \\ of an octopus-inspired robot}%

\author{G. D. Weymouth$^1$, V. Subramaniam$^2$ and M. S. Triantafyllou$^3$}
\address{$^1$ Southampton Marine and Maritime Institute, University of Southampton, 
Southampton SO17 1BJ, United Kingdom.}
\address{$^2$ Centre for Environmental Sensing and Modeling, 
Singapore MIT Alliance for Research and Technology, Singapore 138062, Singapore}
\address{$^3$ Center for Ocean Engineering, Massachusetts Institute of Technology, 
Cambridge MA 02139, United States.}
\ead{G.D.Weymouth@soton.ac.uk}


\begin{abstract}
We design and test an octopus-inspired flexible hull robot that demonstrates outstanding fast-starting performance. The robot is hyper-inflated with water, and then rapidly deflates to expel the fluid so as to power the escape maneuver. Using this robot we verify for the first time in laboratory testing that rapid size-change can substantially reduce separation in bluff bodies traveling several body lengths, and recover fluid energy which can be employed to improve the propulsive performance. The robot is found to experience speeds over ten body lengths per second, exceeding that of a similarly propelled optimally streamlined rigid rocket. The peak net thrust force on the robot is more than 2.6 times that on an optimal rigid body performing the same maneuver, experimentally demonstrating large energy recovery and enabling acceleration greater than 14 body lengths per second squared. Finally, over 53\% of the available energy is converted into payload kinetic energy, a performance that exceeds the estimated energy conversion efficiency of fast-starting fish. The Reynolds number based on final speed and robot length is $Re \approx 700,000$.  We use the experimental data to establish a fundamental deflation scaling parameter $\sigma^*$ which characterizes the mechanisms of flow control via shape change. Based on this scaling parameter, we find that the fast-starting performance improves with increasing size.
\end{abstract}

\noindent{\it Drag reduction, energy efficiency, high-speed maneuvers, shape change \/} \\
\submitto{\BB}
\maketitle
\ioptwocol  
\section{Introduction}

The cephalopods, such as the octopus, cuttle fish, and squid, undergo
large-scale body and volume changes during escape maneuvers. For example, the
octopus first hyper-inflates its mantle cavity filling
it with water, which it then rapidly expels in the form of a
propelling jet \cite{Huffard2006,Wells1990,Packard1969}, accelerating it up to high speed (\fref{fig:octobot}a). Expelling a large mass
of water at lower speed imparts maximum momentum for a given energy expenditure
since momentum scales linearly with speed and mass, while energy scales
linearly with mass but quadratically with speed.  Hence the inflation
before the onset of the maneuver must be large, increasing the body's
lateral dimension substantially. As a result, the normally streamlined
mantle becomes quite bluff. Flow around a similarly shaped rigid body
geometry would incur large energy penalties in the form of flow
separation and added mass resistance.  However, the flexible, rapidly
deflating mantle completely alters the dynamics of the flow, inducing
unsteady flow control mechanisms such as separation elimination and fluid energy recovery.

Unsteady flow control is an active area of research, especially in
relation to animal behavior
\cite{Daniel1984,Kanso2005,Childress2011,Moored2012},
and unsteady propulsion mechanisms involving rapid shape-change offer
intriguing possibilities to overcome the maneuvering limitations of
standard rigid underwater vehicles. The importance of shape-change is exemplified by considering the fluid kinetic energy associated with the body's added mass. While
added mass is fixed for a rigid vessel and the added mass force opposes acceleration,
a shrinking body has a \textit{diminishing} added mass, and can therefore see a 
net flow of energy \textit{into} the body. To
quote from \citeasnoun{Spagnolie2009}, who studied a shape-changing body,
``while a reduced virtual mass gives a reduced acceleration reaction,
a reducing virtual mass can generate a boost in velocity.'' Such a boost is observed in jellyfish even during the `rest' phase of their propulsion cycle, indicating energy recapture \cite{Gemmell2013}. 

This recovery mechanism is especially important to the ultimate speed and efficiency of shrinking bodies performing fast-start maneuvers.
\citeasnoun{Weymouth2013JFM} demonstrated that the irrotational energy initially imparted to the fluid by an accelerating body can be 
recovered by size reduction to increase the propulsive thrust later in the maneuver. In other words, while the added-mass is essentially additional payload for a rigid accelerating body,  a shrinking body effectively turns the added mass into additional propellant. A remarkable consequence is that 
a shrinking rocket in a heavy inviscid fluid accelerates up to speeds greater than it could 
achieve in the vacuum of space.

The amount of irrotational kinetic energy recovery that can be achieved in a viscous fluid is determined by the evolution of the boundary layer vorticity at the external surface of the body as it undergoes large deformations. If this vorticity is shed into the wake the kinetic energy will be lost to the fluid instead of being recovered by the body, hence severely reducing the escape speed of the animal.  \citeasnoun{Weymouth2012JFM} used viscous simulations to
demonstrate that if a body very rapidly undergoes a prescribed shape change, as in the case of the octopus, then the boundary layer
does not shed. Instead, it remains mostly attached to the body, and
the total circulation decreases through vorticity annihilation.

A simple kinematic parameter, the shape-change number $\Xi$, can be used to characterize the relative speed of the size change
\begin{equation}\label{eq:xi}
\Xi=\frac{W}{U}
\end{equation}
where $W$ is the rate at which the cross-stream diameter of the body decreases and $U$ is the forward velocity of the body. \citeasnoun{Weymouth2013JFM} showed that an initially spherical shrinking body of length $L$ that accelerates from rest with constant acceleration $a$ has an ultimate shape change number of $\Xi=W^2/aL$, and that the larger the value of $\Xi$ the better the performance of the shrinking body. Specifically, simulations up to Reynolds number $Re=1,000$ showed that moderate separation reduction and energy recovery required $\Xi > 1/32$.

However, the purely kinematic parameter $\Xi$ is incapable of fully characterizing the physics of the boundary layer evolution. For a self-propelled body
simulated in \citeasnoun{Weymouth2013JFM} with a maximum Reynolds number $Re= UL/\nu=20,000$ but a
small value of $\Xi \approx 1/44$, it was found that
separation was practically absent and energy recovery remarkably high.
This demonstrated that Reynolds number is also a significant parameter
but the specific dependence was not resolved. Limitations in the simulation also 
left unresolved questions concerning the energetics and the feasibility of physically implementing 
a propulsion-system based on rapid size-change.

In this paper we apply and extend the concepts derived in previous work to
design and test a deflatable flexible hull robot inspired by the octopus. The initially bluff robot demonstrates remarkable fast-starting performance through
separation elimination and flow energy recovery.  We use the results
to investigate the principal physical mechanisms involved and derive a single
fundamental parameter capable of characterizing the flow around deflating bodies.

\section{Design of an octopus-inspired robot}

\begin{figure}
	\includegraphics[width=0.4\textwidth]{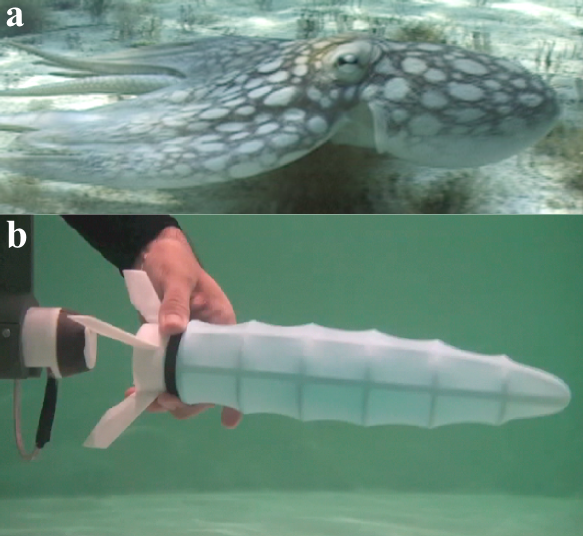}
	\caption{Biological and biologically-inspired jet-propelled escape. (a)~Octopus 
		  using jet propulsion to escape from threat at high
          speed. Image captured from video courtesy of Dr.~Roger
          Hanlon of the Marine Biology Laboratory, Woods Hole. (b)~This 
          propulsion mode inspired the design of a simple jetting
          robot using a 3D printed polycarbonate model and covered by
          an elastic membrane made of synthetic
          rubber. \label{fig:octobot}}
\end{figure}

To test the ability of shape-change to reduce separation and recover fluid energy in experiments required the design of the novel underwater robot shown in \fref{fig:octobot}b. The octopus-inspired robot consists of a rigid neutrally buoyant skeleton with an elastic membrane stretched around it to form the outer hull. As with the mantle of the octopus, this membrane can be inflated, giving it an initially bluff shape and storing sufficient
energy to power its escape. The fully deflated hull shape is approximately a 5:1 ellipsoid, and is sufficiently streamlined to allow the body to coast dozens of body lengths. 

\begin{figure}
	\includegraphics[width=0.4\textwidth,trim=0 0 0 5cm,clip=true]{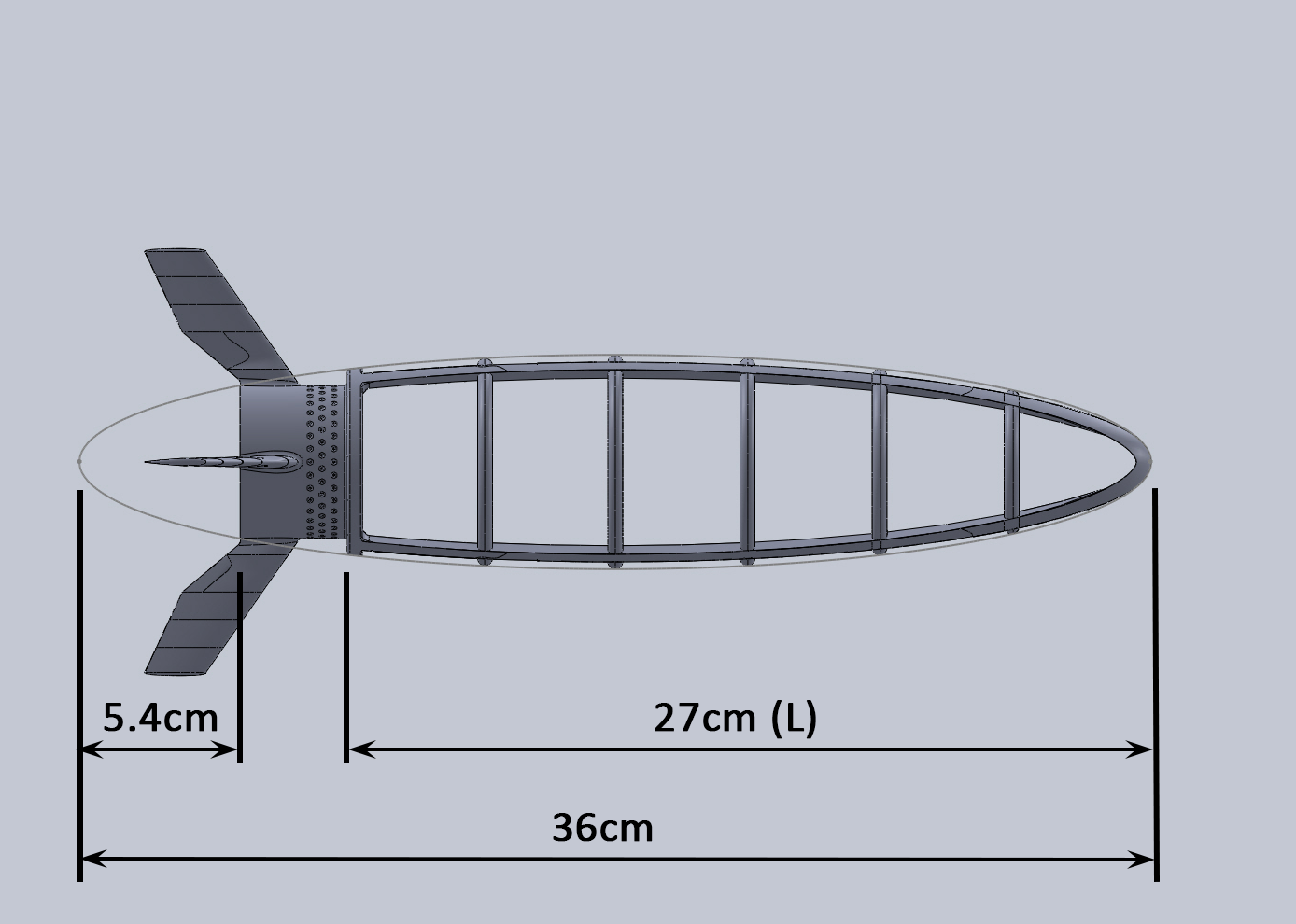}
	\caption{Dimensions and shape of internal skeleton of the
          robot, 3D printed using polycarbonate. The length of the
          membrane-covered section is $L=27\,cm$, and a 5:1 ellipse is drawn on the background for reference.\label{fig:octobot_length}}
\end{figure}

The rigid skeleton was 3D-printed in a single piece out of polycarbonate
(\fref{fig:octobot_length}). This structure has large openings to 
reduce frictional losses from the fluid as it is pushed through the 
robot when the membrane deflates. The length of the membrane-covered
section is $L=27\,cm$ to match the inflated diameter of the commercially available synthetic rubber balloons. As the skeleton is neutrally buoyant and the membrane is filled with water, the robot is neutrally buoyant throughout the maneuver. The volume of the robot when fully deflated is $1030\,cm^3$, and so the `payload' mass accelerated by the maneuver is $m_f=1.03\,kg$. 

\begin{figure}
	\includegraphics[width=0.45\textwidth]{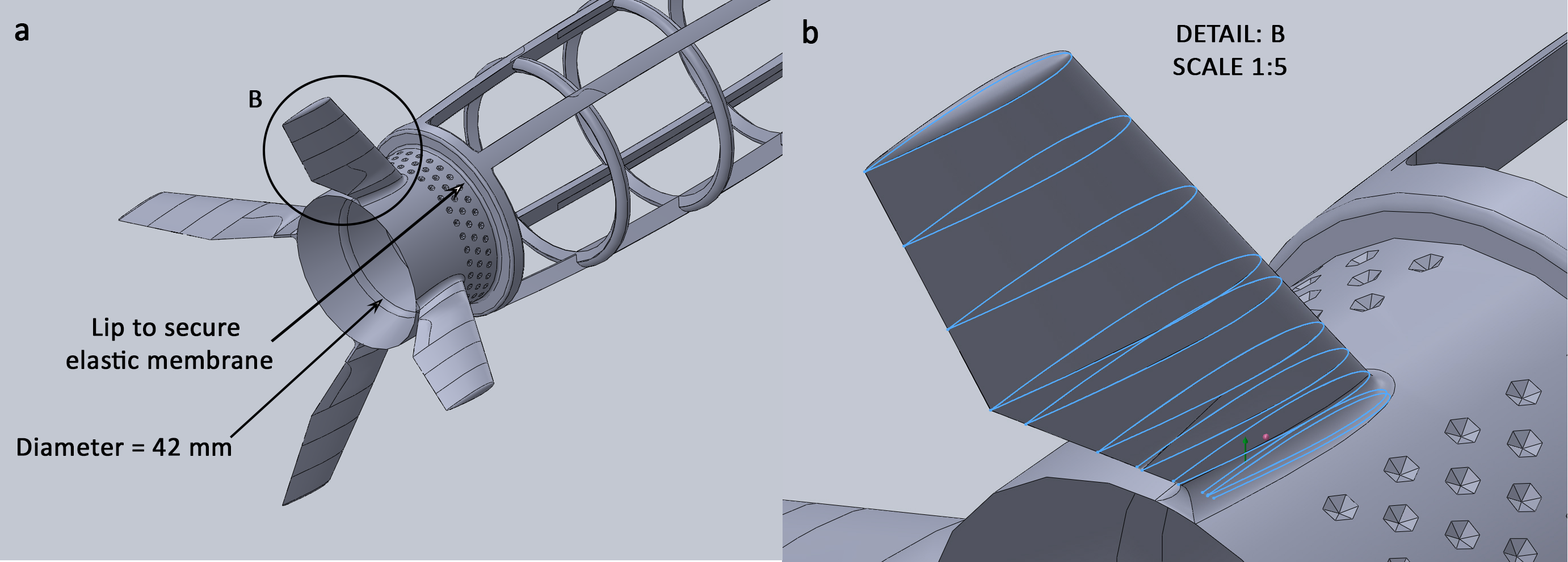}
	\caption{(a) Zoomed in isometric view of the tail of the robot showing the large openings and the dimensions and shape of the jet nozzle. The membrane is pretensioned along the length and fastened to the skeleton at the rim indicated. (b) Fins with a NACA 0012 cross-sections are used acheive stable open-water motion of the robot.\label{fig:octobot_funnel}}
\end{figure}

The tail of the skeleton is fitted with a convex jet nozzle and a set of four NACA 0012 fins for directional stability instead of the trailing arms of the octopus, hence avoiding interactions with the propelling jet (\fref{fig:octobot_funnel}). The aperture area $A_J= 15\,cm^2$ was chosen as a compromise between the need to have a large aperture to increase the flow rate and the requirement to have a small area that fits the ultimate (deflated) shape.

\section{Open-water testing methodology}
 
\begin{figure}
	\includegraphics[width=0.35\textwidth,trim=0 0 0 2cm,clip=true]{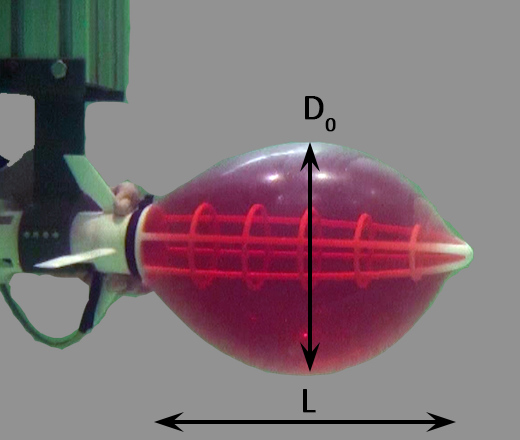}
	\caption{The robot is filled with pressurized water (died red in this image for contrast) from a
          rigid mount, hyper-inflating the balloon to take on a bluff
          body shape. The initial inflated diameter is $0.6\,L$ for every test.\label{fig:inflate}}
\end{figure}

The simple but effective design above does not require any moving parts or
energy storage other than the stretching membrane and enables self-propelled open water 
fast-start tests to be performed. In these tests, the robot is slowly inflated by filling it with
pressurized water from a rigid mount which was fitted with a pressure sensor (\fref{fig:inflate}). The robot was filled until the diameter of the membrane reached $0.6\,L$ giving a pressure difference inside and outside the membrane of $4.45\,kPa$. In this condition, the membrane has a surface area of $0.1\,m^2$. The strain in the membrane is highly localized, but the overall strain relative to the fully deflated shape is $\epsilon=0.94$. Thus the effective modulus is only $4.77\,kPa$. Inflation past this point typically led to membrane failure. 

\subsection{Measurement and image processing method}

\begin{figure}
	\includegraphics[width=0.45\textwidth]{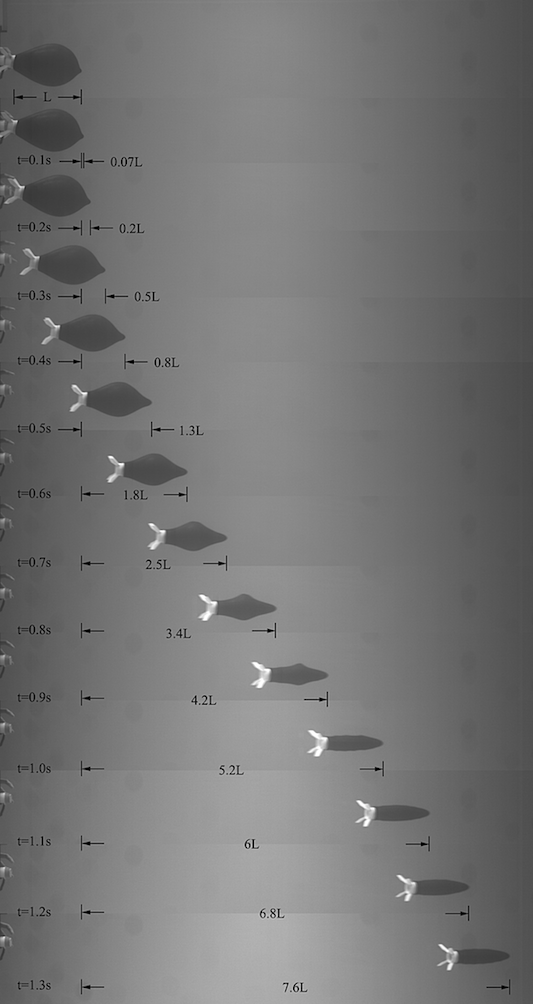}
	\caption{Filmstrip of the octopus-inspired robot as it jets and shrinks in the self-propelled fast-start maneuver. The time delay between images is 0.1 seconds. \label{fig:filmstrip}}
\end{figure}

Once inflated, the robot is released from the mount allowing it to accelerate forward in open water under its own power.
The resulting fast-start maneuver performance is measured using high-speed cameras at
150 frames/second. \Fref{fig:filmstrip} shows the rapid acceleration and deflation 
of the shrinking robot from a self-propelled run.

\begin{figure*}
	\includegraphics[width=0.8\textwidth]{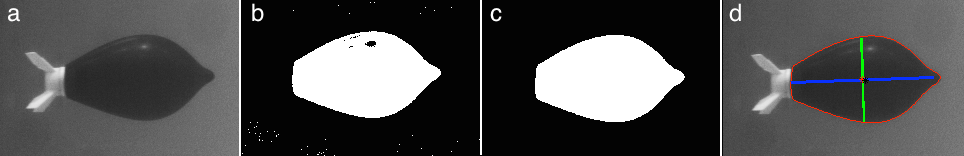}
	\caption{Each frame of the high speed camera data (a) was processed using thresholding (b), and edge dection algorithms (c), to determine the instantaneous centroid and outline of the robot (d), which thereby determine the robot kinematics and dynamics. \label{fig:video-frame}}
\end{figure*}

Each frame of the high-speed video is analyzed separately, and a sample of the process is depicted in \fref{fig:video-frame}. The algorithm applies a threshold to the image followed by edge detection and morphological operations. The end result is an image with the robot isolated from the background. The length scale in each image is calibrated based on the known dimensions of the rigid skeleton. From this the position of the membrane outline is established to a resolution of $\pm 0.9\,mm$, corresponding to the average size of one pixel in the image. The outline is numerically integrated to give the instantaneous volume $V$ and center of mass enclosed by the flexible hull assuming the membrane to be axisymmetric about the robot's major axis. Differentiating the location of the center of mass between snapshots gives the instantaneous velocity $U$. We find that the measurement error of the membrane tracking method is averaged out in the volume integration, resulting in consistent and noise free measurements of the robot volume and mass. In contrast, the finite differencing introduces numerical noise, which we filter by fitting a smoothing spline to data from four test runs.

\section{Results of self-propelled robot tests}

\begin{figure*}
	\includegraphics[width=\textwidth,trim=1mm 0 3mm 0,clip=true]{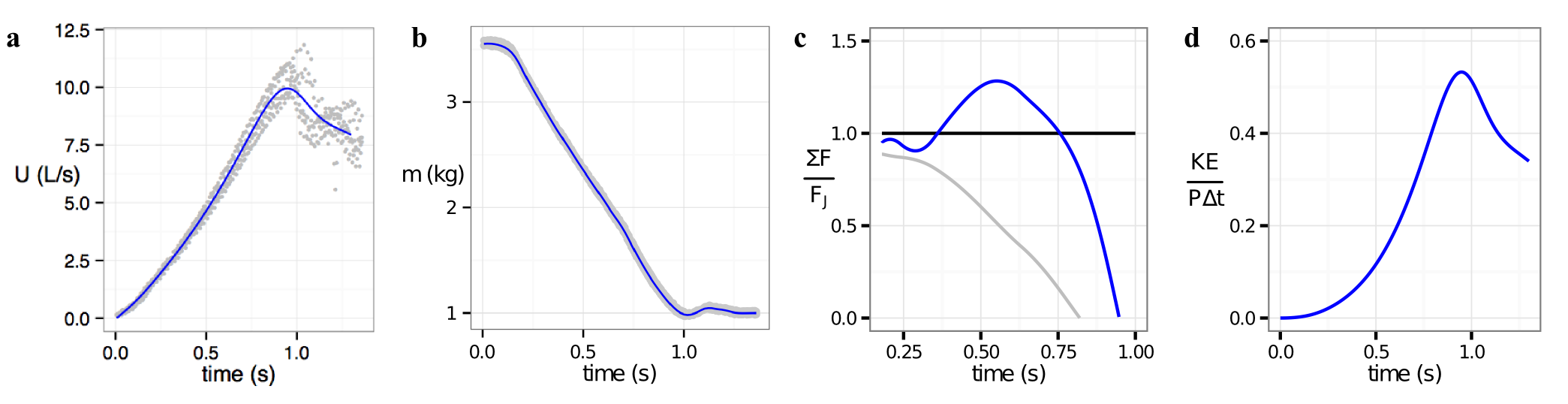}
	\caption{Data analysis from four self-propelled runs.
          (a) Velocity measurements (gray points) show that the
          robot achieves peak velocity $U$ greater than $10L/s$
           in less than a second. (b) The mass $m$ of the robot (gray points) decreases as the body shrinks, ejecting fluid at 
          a rate $\dot m = -3\,kg/s$. The blue lines in (a,b) are the average performance values using a smoothing spline. (c) The peak net thrust force
          $\Sigma F$ is 30\% greater than the thrust provided by the
          jet $F_J$. The gray line shows a conservative estimate of
          the net force on a streamlined body with the shape of a 5:1
          spheroid ($m_a=m/10$, $C_D=0.05$ \cite{Hoerner1965}) under
          the same conditions. (d) The high thrust enables 53\% of the
          integrated jet power $P\Delta t$ to be converted into
          payload kinetic energy $KE$.\label{fig:results}}
\end{figure*}
 
Qualitatively we see from, \fref{fig:filmstrip} that the robot moves slowly in the first $0.5\,s$, but the boost in speed achieved at the end of the maneuver is clearly visible, with the robot displacing the membrane length $L$ between the frames at $0.9$ and $1.0\,s$. The final streamlined shape allows the robot to maintain fairly high speed, similar to the deflated form of escaping cephalopods.
 
The velocity history presented in \fref{fig:results}a quantifies these observations. The robot achieves peak velocity $U$ above $10\,L/s$
(body lengths per second) or $2.7m/s$ at $t=0.95\,s$, after which the robot slows down as it coasts. The acceleration is fairly constant during the jetting period, with a peak value of $14\,L/s^2$ or $3.8\,m/s^2$. The peak velocity gives a maximum
Reynolds number of 729,000, well above the value of 20,000 explored
numerically, and consistent with the values experienced by fast moving
marine animals \cite{Daniel1984}. The value of $\Xi=W_{ave}/U_{max}$ is approximately 1/24.

\Fref{fig:results}b shows that the mass $m$ of the robot decreases from the initial size of $3.65\,kg$ down to the payload mass when the membrane comes in contact with the rigid skeleton. At the end of the maneuver, the membrane is less stiff and in contact with the struts of the rigid structure leading to a ``starved horse'' effect and the observed dip and recovery in the robot mass at $t=1\,s$. The mass decreases at a rate near $3\,kg/s$ through most of the jetting period. 

\subsection{Force analysis and efficiency}

To extend this analysis, the filtered values of $U$ and $m$ are used to
compute the net force on the robot $\Sigma F= m \dot U$ and
the jet force $F_J=\dot m U_J$, where $U_J=\tfrac 43 \dot V/A_J$ is the
jet momentum velocity. $F_J$ is a conservative estimate of the jet momentum flux as it assumes a parabolic profile, a uniform profile produces 33\% less thrust for the same mass flow rate, but it neglects any additional forces on the nozzle. Additional thrust can be produced by a pressure impulse in jet flow \cite{Krueger2005,Krieg2013}, but these forces are strongest for pulsed jets and were not observed to be significant in the previous numerical work comparing shrinking and rigid bodies \cite{Weymouth2013JFM}. The proximity of the filling mount early in the maneuver does induce a pressure force on the nozzle, and so the $F_J$ model is only used after $t=0.2\,s$.

\Fref{fig:results}c shows that the instantaneous net force on the body is remarkably high, peaking at 130\% of the force provided by the jet. For comparison, \fref{fig:results}c also shows a conservative estimate of the forces that would be experienced by a rigid 5:1 prolate body made to follow the same kinematics experienced by the robot. A 5:1 ellipsoid is a nearly optimal streamlined form for steady forward motion \cite{Hoerner1965} but the drag and reactive forces on a rigid ellipsoid would still reduce the net force by 50\% at $t=0.6\,s$. Therefore, the net force on the shrinking robot is 2.6 times that of an optimal rigid body at $t=0.6\,s$, and the relative improvement is even greater later in the maneuver.

Finally, we compute the kinetic energy of the payload $KE=\frac
12 m_f U^2$ and the integrated power (total work) supplied by the
jet $P\Delta t=\frac 12\int\dot m U_J^2 dt$. We use the final payload mass $m_f=1.03\,kg$ instead of the instantaneous mass to avoid crediting the robot for accelerating propellant that it ejects later in the maneuver. The ratio the payload kinetic energy and the work
done by the jet gives the hydrodynamic energy efficiency of the robot
in accelerating its payload up to speed. The large energy recovery
powered thrust of the shrinking body enables a correspondingly large
energy efficiency with a peak of 53\% during the fast-start maneuver.

\section{Physical mechanisms exploited by size-reducing bodies}

We next address the physical mechanisms of separation reduction and
energy recovery to establish general rules for their application to
other situations of shape-changing bodies. The key factors are the
strong velocity component normal to the body surface and the large
pressure gradients generated by the rapid shrinking. 

A simplified potential flow model of the flow around translating and shrinking
sphere demonstrates these factors. The potential in spherical
coordinates ($r,\theta,\psi$) is:
\begin{equation}\label{eq:potential}
\phi = -U\left(r+\frac{R^3}{2r^2}\right)\cos\theta-\frac {WR^2}{2r}
\end{equation}
where $R$ is the instantaneous radius of the sphere, $W/2=\dot R$ is the rate of change of the radius. Taking the gradient of the unsteady potential at $r=R$ gives the velocity on the surface of the sphere
\begin{align}
u_r &= \frac{\partial\phi}{\partial r} = U\cos\theta + \frac W2 \label{eq:normal}\\
u_\theta &= \frac 1r\frac{\partial\phi}{\partial \theta} = \frac 1 2 U \sin\theta
\end{align}
showing that the radial component of velocity matches the required normal condition on the shrinking sphere.

The unsteady Bernoulli equation gives the pressure on the surface of the sphere as
\begin{equation}\label{eq:press}
p = \frac 9{16}\rho U^2\cos 2\theta+\frac 32\rho\left(\frac{UW}2+Ra\right)\cos\theta
\end{equation}
where $a=\dot U$ as before and we have dropped the terms that are uniform on the surface ($W^2$,$\dot WR$) as they do not contribute to the tangential pressure gradient or to the net force.
The first term in \eref{eq:press} is the pressure due to the
forward speed of the body, proportional to $\rho U^2$. The second term
is due to the rate of change of the body's size and speed, 
proportional to $\rho U W$ and $\rho R a$, respectively. From \eref{eq:press} we see that $\Xi=W/U$ expresses the magnitude of the pressure induced by shrinking and translating relative to the pressure induced purely by translation.

Large negative values of $W$ cause low pressure at the front of the body and high pressure at
the back. The integrated effect of this pressure is a forward thrust force completely separate from any decrease in the quasi-steady drag caused by the reduced cross
stream area. As discussed in the introduction, this thrust force is the result of the 
transfers fluid kinetic energy $T=\frac 12 m_a U^2$ back into the
body, as the added mass $m_a$ decreases with the size $R$. The
magnitude of this force is:
\begin{equation}\label{eq:thrust}
F = -\frac d{dt}(m_a U) = -\dot m_aU-m_a a
\end{equation} 
where $\dot m_a$ depends linearly on the shrinking rate $W$. \Eref{eq:thrust} demonstrates that when the body is bluff and moving slowly early in the maneuver, the force is negative as energy is transferred into the fluid. However, later in the maneuver this energy can be recovered by a body that is shrinking and moving quickly. This accounts for the slow start but rapid finish observed in the octopus robot results.

\subsection{Separation reduction at high Reynolds number}

The recovery of fluid kinetic energy is perfect within irrotational
flow, but in viscous fluids the added mass-induced thrust depends
critically on simultaneous separation reduction to avoid
irrecoverable energy loss to the wake in the form of large vortical
structures. Numerical simulations of shrinking circular cylinders in
\cite{Weymouth2012JFM} demonstrated that the negative pressure
gradient due to $W$ in \eref{eq:press} generates a sheet of
opposite sign vorticity on the surface of the body, starting from the
rear and moving up towards the 90 degree point. At the same time, the
normal velocity on the wall induced by shrinking advects
the previously formed boundary layer vorticity towards the body
surface, causing it to come in contact with the newly developed layer
of opposite-sign vorticity, resulting in partial vorticity
annihilation.  This is similar to the vorticity annihilation noted by
\cite{Kambe1986} in the case of two shear layers of opposite vorticity
forced to collide.

This mechanism of vorticity control depends on $W$ overcoming the
rate of diffusion and separation in the boundary layer. To estimate
the required rate of shape change it is insightful to compare a
shrinking deformable body to the application of suction on a rigid
body boundary layer \cite{Choi2008}.  Both processes induce
velocity normal to the body surface (as seen in \eref{eq:normal}), although shrinking induces it without
actual mass-flow through the membrane.  As shown in the study of a
porous circular cylinder of diameter $D$ placed in cross-flow $U$
\cite{Pankhurst1953}, employing a uniform suction with velocity $w_o$
causes delayed separation and decreased drag when the suction
coefficient, $C_q=\pi D w_o/(U D)$, exceeds a theoretical value
estimated by Prandtl as $C_q=4.35 \pi/\sqrt{Re}$ and by
\citeasnoun{Preston1948} as $C_q=3.214 \pi/\sqrt{Re}$. For example, for $C_q
\sqrt{Re} = 14$ the drag coefficient is reduced from a value of 0.90
without suction to a value of 0.69; and for $C_q \sqrt{Re} = 42$ the
drag coefficient is measured to be 0.01.

Returning to the deflating robot problem, the analogy with applying
suction can be used to define equivalent controlling parameters,
particularly in view of the circulation reduction noted above for the
case of a shrinking cylinder.  Through the use of the Mangler
transform \cite{Mangler1948,Schlichting2000}, three-dimensional axisymmetric boundary layers can be
reduced to planar ones hence we define a {\em deflation scaling parameter}, $\sigma^*$, which is
analogous to the suction scaling parameter of a cylinder:
\begin{equation}\label{eq:sigma}
\sigma^* = \frac{\dot V}{AU}\sqrt{Re}
\end{equation}
where $A$ is the frontal area, and $\dot V$ is the rate of change of the volume of the body, which is proportional to the average normal velocity on the surface, and therefore typically to $W$. The deflation scaling parameter
$\sigma^*$ therefore provides an instantaneous value that accounts for the
effects of the shape-change number as well as the Reynolds number, hence replacing
the parameter $\Xi$ in predicting separation control. We
estimate the theoretical threshold value of $\sigma^*$ for separation
delay on a sphere as:
\begin{equation}\label{eq:sigma_thresh}
\sigma^* ~>~ 2.41 \pi
\end{equation}
following the methodology of \citeasnoun{Preston1948}.

We can test the predictive power of this parameter using the previous and current results. 
In \citeasnoun{Weymouth2013JFM} $\Xi=1/32$ was
found to result in moderate separation because the average $Re$ during 
the maneuver was 500, giving a low value of $\sigma^*=1.85$. In contrast, in the
simulation of the rocket fast-start maneuver we observed good boundary
layer control and energy recovery despite the small value of
$\Xi\approx 1/44$ because the higher $Re$ (peaking at $20,000$) results in 
$\sigma^*>9$ throughout the maneuver.

The value of $\sigma^*$ for the deflating robot is adequate to effectively control the flow and recover the fluid kinetic energy. The minimum value of $\sigma^*$ occurs at around $t=0.5\,s$, while the robot is still fairly bluff (the area is $A=170\,cm^2$) but has accelerated up to fairly high speed ($U=135\,cm/s$). The deflation rate of the volume $\dot V=3000\,cm^3/s$ therefore gives $\sigma^*_{min}=77$ for the robot escape maneuver. This minimum value is well above the threshold, enabling separation control and added-mass energy recovery and leading to the remarkably high measured net force shown in \fref{fig:results}c.

\section{Conclusions}

We designed and tested a flexible hull inflatable robot that exhibits
outstanding fast-starting performance, emulating the function of an
escaping octopus for the amount of power available, reaching final
speeds in excess of 10 body lengths per second in less than a second,
and converting 53\% of the energy to kinetic energy.

We note that the energetics of the self-propelled robot are
particularly important, as powering unsteady maneuvers can be the
limiting factor in the operational life of self-powered underwater
vehicles. The overall efficiency of energy conversion at 53\% is at
least as high as the best in fast-starting fish \cite{Frith1995}; and
the robot achieves speeds matching those of fast squid
\cite{Neumeister2000}. While the top accelerations for fast-starting
fish are reported at 40 to 120 $m/s^2$, they must be scaled with the
amount of energy available for the maneuver
\cite{Gazzola2012,Domenici1997}. In our experiments the extensibility
and ultimate strength of the commercially available membrane we used
placed an upper limit on the initial internal pressure, $\delta p=4.45\,kPa$,
and therefore on the initial energy of the robot and the force from
the propelling jet. However, the deflation rate $W$ determines how
efficiently the energy is recaptured, and this is largely determined by 
the geometry for freely accelerating bodies.
In addition, the form of the parameter $\sigma^*$ shows that
as the Reynolds number increases the required threshold deflation rate
decreases, making flow control even more feasible. Therefore, for a
given robot design and higher available energy, the efficiency is
expected to remain near the (high) level found in these experiments,
and an increase in membrane stiffness will enable man-made vehicles to
rival the acceleration of their biological inspirations.

\subsection*{Acknowledgments}
The authors wish to acknowledge support from the Singapore-MIT
Alliance for Research and Technology through the CENSAM program, and
from the MIT Sea Grant program. We also thank Joshua C Born for
assisting with data acquisition during his internship and Roger Hanlon 
for valuable discussions and videos regarding cephalopod propulsion.

\section*{References}

\end{document}